\documentclass[%
 reprint,
 superscriptaddress,
 nobibnotes,
 amsmath,
 amssymb,
 aps,
 prm,
 floatfix,
]{revtex4-2}

\usepackage[caption=false]{subfig}
\newcommand{\phantomsubfloat}[1]{
    {
        \captionsetup[subfloat]{farskip=0pt,captionskip=0pt}
        \captionsetup[subfigure]{labelformat=empty}
        \subfloat{#1}
    }%
}

\usepackage[T1]{fontenc}
\usepackage[dvipsnames]{xcolor}
\usepackage[normalem]{ulem}
\usepackage{amsfonts}
\usepackage{graphicx}
\usepackage{dcolumn}
\usepackage{bm}
\usepackage{hyperref}

\hypersetup{
    hidelinks,
    colorlinks=true,
    linkcolor=blue,
    filecolor=blue,      
    urlcolor=blue,
    citecolor=blue
}

\begin{document}

\title{Equivariant graph neural network interatomic potential for \\ Green-Kubo thermal conductivity in phase change materials}

\author{Sung-Ho Lee}
\affiliation{Univ. Grenoble Alpes, CEA, Leti, F-38000, Grenoble, France}

\author{Jing Li}%
\affiliation{Univ. Grenoble Alpes, CEA, Leti, F-38000, Grenoble, France}

\author{Valerio Olevano}
\affiliation{
Institut Néel, CNRS \& Univ. Grenoble Alpes, F-38042, Grenoble, France
}%

\author{Benoit Sklénard}
 \email{benoit.sklenard@cea.fr}
\affiliation{Univ. Grenoble Alpes, CEA, Leti, F-38000, Grenoble, France}

\date{\today}

\begin{abstract}
Thermal conductivity is a fundamental material property that plays an essential role in technology, but its accurate evaluation presents a challenge for theory.
In this work, we demonstrate the application of $E(3)$-equivariant neutral network interatomic potentials within Green-Kubo formalism to determine the lattice thermal conductivity in amorphous and crystalline materials.
We apply this method to study the thermal conductivity of germanium telluride (GeTe) as a prototypical phase change material.
A single deep learning interatomic potential is able to describe the phase transitions between the amorphous, rhombohedral and cubic phases, with critical temperatures in good agreement with experiments.
Furthermore, this approach accurately captures the pronounced anharmonicity that is present in GeTe, enabling precise calculations of the thermal conductivity. In contrast, the Boltzmann transport equation including only three-phonon processes tends to overestimate the thermal conductivity by approximately a factor of 2 in the crystalline phases.
\end{abstract}

\maketitle

\section{Introduction}

Thermal conductivity, as an intrinsic material property, holds significant implications in technology as it plays a crucial role in determining the thermal management of electronic devices~\cite{krinner-2019,zhang-2022a} and serves as a key parameter in thermoelectric device performance~\cite{zhang-2020,gutierrez-2020}. Heat transport in semiconductors and insulators is primarily governed by lattice vibrations, i.e., phonons. Substantial efforts have been directed towards precise calculations of lattice thermal conductivities from a microscopic perspective.
The main methods for calculating lattice thermal conductivity include the Boltzmann transport equation (BTE)~\cite{broido-2007,sklenard-2021,li-2022}, nonequilibrium Green's function (NEGF) theory~\cite{mingo-2006,li-2009,guo-2020}, nonequilibrium molecular dynamics (NEMD)~\cite{muller-plathe-1997} and the Green-Kubo (GK) formula~\cite{fan-2015,isaeva-2019,zhang-2022b}.
The BTE assesses the response of phonon occupation to a temperature gradient, typically including three-phonon scattering processes, which limits its applicability to weakly anharmonic crystalline materials. In contrast, NEGF treats phonons quantum mechanically, considering interface scatterings and phonon anharmonicity through self-energies, but it comes with a computational cost~\cite{mingo-2006}.
The NEMD method is based on a direct simulation of stationary heat fluxes in a nonequilibrium state and naturally include anharmonic effects.
Finally, GK provides the lattice thermal conductivity from the autocorrelation of heat fluxes usually computed in an equilibrium molecular dynamics (MD) simulation, accounting for anharmonic effects to all orders~\cite{tristant-2019}. 
Furthermore, recent developments extend GK to low temperatures~\cite{isaeva-2019,zhang-2022b}, which makes it a robust approach for a wide range of temperatures and materials. GK theory provides a unified approach to compute the lattice thermal conductivity in ordered and disordered solids. For harmonic amorphous systems, thermal transport can be described by the Allen and Feldman (AF) theory~\cite{allen-1993}. However, it has been shown that the AF theory may be inadequate when anharmonic effects become important~\cite{shenogin-2009,xueyan-2022}.

The MD simulation in the GK approach requires a relatively long simulation time (up to a few nanoseconds) for sufficient statistical sampling and an accurate description of interactions among atoms. 
Such long simulation times are affordable for MD with empirical force fields, but at the price of reduced accuracy and universality. 
\textit{Ab~initio} MD has better accuracy but is too computationally expensive for large systems or long MD simulations.
Extrapolation schemes have been proposed~\cite{carbogno-2017} to reduce the computational cost, but they are unsuitable for disordered solids. 

Over the past few years, machine learning has emerged as a practical alternative for tasks where \textit{ab~initio} methods encountered difficulties. Notably, machine learning interatomic potentials (MLIP) have proven successful in rapidly predicting energies, forces, and stress tensors with an accuracy comparable to first-principle methods. Particularly in thermal transport GK calculations, MLIPs have been employed using descriptor-based approaches, such as Behler-Parrinello neural networks or kernel-based methods~\cite{verdi-2021,korotaev-2019,sosso-2012}.
Recently, neural network interatomic potentials utilizing message passing architectures~\cite{schutt-2017,batzner-2022,schutt-2021,brandstetter-2022} have been proposed as an alternative to hand-crafted descriptors. In this approach, structures are encoded as graphs, where atoms are depicted as nodes connected by edges.
In initial models, the information at the nodes and edges was made \textit{invariant} with respect to the Euclidean group $E(3)$ (i.e., the group of translations, rotations, and inversions in Euclidean space), and the atomic representations were limited to scalar interatomic distances~\cite{schutt-2017}. Such models have since been largely superseded by architectures built on convolution operations that are \textit{equivariant} with respect to the $E(3)$ group. In equivariant approaches, isometric transformations on the relative atomic displacement vector inputs are propagated through the network to correspondingly transform the outputs. Equivariant MLIPs have been shown to achieve substantially improved data efficiency and unprecedented accuracy compared to their invariant counterparts~\cite{batzner-2022,schutt-2021,brandstetter-2022}. However, in such message passing architectures, many-body interactions are captured by iteratively propagating information along the graph at each layer in the network. This has the effect of extending the local receptive field of an atom to significantly beyond the cutoff radius, which renders parallelization impractical~\cite{musaelian-2023}. Recently, a strictly local equivariant neural network potential has been proposed to address this drawback~\cite{musaelian-2023}. In this architecture, information is stored as a per-pair quantity, and instead of nodes exchanging information with its neighbours via edges, a convolution operation acts on the cutoff sphere in the form of a set of invariant (scalar) latent features and a set of equivariant (tensor) latent features that interact at each layer.

In this work, we employ the strictly local $E(3)$-equivariant neural network potential from Ref.~\cite{musaelian-2023} to compute the temperature-dependent thermal conductivity of germanium telluride (GeTe) in various phases using GK theory. GeTe is a chalcogenide material employed in many technological applications, such as phase change nonvolatile memory storage~\cite{lankhorst-2005,jeong-2021}, thermoelectricity~\cite{zhang-2020,zhang-2021,jiang-2022} and spintronics~\cite{picozzi-2014,rinaldi-2016,varotto-2021}. It undergoes a ferroelectric phase transition from the low temperature rhombohedral $\alpha$-GeTe (spacegroup $R3m$) to the cubic $\beta$-GeTe (spacegroup $Fm\bar{3}m$) at a Curie temperature of $T_c \approx650-700$~K~\cite{chatterji-2015, sist-2018, chattopadhyay-1987}. Amorphous GeTe also plays an important role in technological applications such as phase change memories. Hence, GeTe serves as a prototype material to demonstrate the universality of the GK method combined with equivariant neural network interatomic potentials. This approach effectively captures strong anharmonicity and materials in diverse crystalline phases or amorphous states for the calculation of lattice thermal conductivity.

\section{Methodology}

The thermal conductivity tensor within GK theory is defined as
\begin{equation}
\label{eq:gk_kappa}
    \kappa_{\alpha\beta}(T) = \dfrac{1}{k_\mathrm{B} T^2 V} \lim_{\tau\to\infty} \int_{0}^{\tau} dt \, \langle j_\alpha(t) \cdot j_\beta(0) \rangle_{T}
    ,
\end{equation}
where $k_\mathrm{B}$ is the Boltzmann constant, $T$ the temperature, $V$ the volume, $j_\alpha(t)$ the $\alpha$-th Cartesian component of the macroscopic heat flux, and $\langle j_\alpha(t) \cdot j_\beta(0) \rangle_{T}$ the heat flux autocorrelation function, with the symbol $\langle \cdot \rangle_T$ denoting the ensemble average over time and over independent MD trajectories. 

The total heat flux of a system of $N$ atoms is given by
\begin{equation}
\label{eq:hf}
\bm{j}(t) = \sum_{i=1}^{N} \dfrac{d}{dt} \left( \bm{r_i} E_i \right)
,
\end{equation}
where $E_i = m_i \bm{v}_i^2 / 2 + U_i$ is the total energy (i.e.\ kinetic and potential energy) of atom $i$ with mass $m_i$, velocity $\bm{v}_i$ and atomic position $\bm{r_i}$. In MLIPs, the partitioning $U = \sum_i U_i $ of the total energy of the system into atomic contributions $U_i$ allows the total heat flux of a periodic system to be expressed as~\cite{fan-2015}
\begin{equation}
\label{eq:heatflux}
\bm{j}(t) = \sum_{i=1}^{N} \bm{v}_i E_i - \sum_{i=1}^{N} \sum_{j \neq i} \bm{r}_{ij} \left( \dfrac{\partial U_i}{\partial \bm{r}_{ij}} \cdot \bm{v}_j \right)
\end{equation}
where the sum over $j$ runs over the atoms that are within the cutoff radius $r_c$ of atom $i$.
We implemented the calculation of Eq.~(\ref{eq:heatflux}) in the \textsc{LAMMPS} code~\cite{thompson_lammps_2022}. The term $\partial U_i / \partial \bm{r}_{ij}$ is obtained by automatic differentiation and was also used for the calculation of the virial tensor~\cite{thompson-2009, fan-2015}, which is required to perform simulations in the isothermal-isobaric (NpT) ensemble.

\section{Machine Learning Interatomic Potential}

\subsection{Training dataset}

To create the reference dataset for training the MLIP, \textit{ab initio} MD simulations based on density functional theory (DFT) were carried out using the \textsc{VASP} code~\cite{kresse-1996,kresse-1999}, with temperatures ranging from 100~K to 2500~K. The generalized gradient approximation of Perdew, Burke, and Ernzerhof (PBE)~\cite{perdew-1996} was used for the exchange-correlation energy and Grimme's D3 dispersion correction~\cite{grimme-2010} was applied. The supercells contained 192 and 216 atoms for the initial rhombohedral and cubic structures, respectively. Then, a total of 6000 structures were extracted from the MD trajectories and recomputed to obtain more accurate energies, forces, and stress tensors. We used an energy cutoff of 400~eV and a $2 \times 2 \times 2$ $k$ mesh to sample the Brillouin zone. The equivariant NN model was trained using the \textsc{Allegro} package~\cite{musaelian-2023}. Details of the training procedure and dataset partitioning are provided in Appendix~\ref{sec:mlip}. The training and test GeTe datasets can be found on Zenodo~\cite{lee-2024}. 
The root mean squared errors (RMSE) and mean absolute errors (MAE) on the predicted energies, forces and stress tensors on the test dataset are $0.90$~meV/atom, $29.87$~meV/$\text{\AA}$, $0.28$~meV/$\text{\AA}^3$ and $1.07$~meV/atom, $42.97$~meV/$\text{\AA}$, $0.37$~meV/$\text{\AA}^3$, respectively.

\begin{figure}[t]
    \centering
    \includegraphics[width=0.95\linewidth]{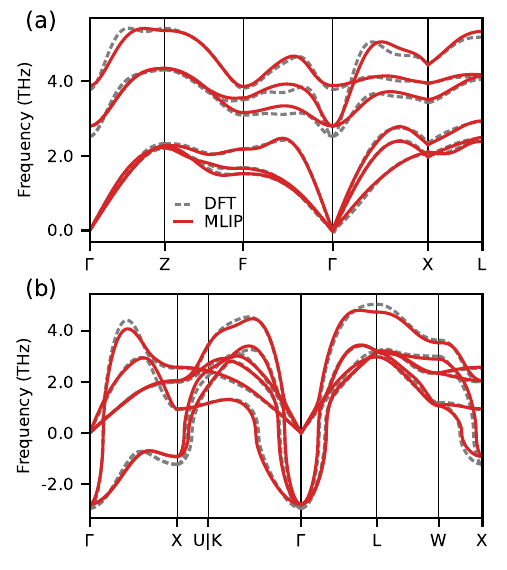}
    \phantomsubfloat{\label{fig:phonon-rhomb}}
    \phantomsubfloat{\label{fig:phonon-cub}}
    \caption{Comparison of phonon dispersions computed with DFT and with the MLIP of \protect\subref{fig:phonon-rhomb} $\alpha$-GeTe and \protect\subref{fig:phonon-cub} $\beta$-GeTe}
    \label{fig:phonon}
\end{figure}

\subsection{Model validation}

To validate the MLIP further, the equilibrium geometries of crystalline GeTe were optimized using the MLIP. For $\alpha$-GeTe, the lattice parameter was $a=4.42$~$\text{\AA}$ and the angle $\alpha=57.13 ^\circ$, closely matching the DFT results ($a=4.41$~$\text{\AA}$ and $\alpha=57.42 ^\circ$). Similarly, for $\beta$-GeTe, the MLIP yields $a = 4.24$~$\text{\AA}$, in excellent agreement with the lattice parameter from DFT of $a=4.23$~$\text{\AA}$.

Moreover, the phonon dispersion obtained from the MLIP is in excellent agreement with DFT for both $\alpha$ and $\beta$-GeTe, as shown in Fig.~\ref{fig:phonon}.
In particular, our model describes optical phonons well, which is usually challenging for MLIPs~\cite{liu-2021,verdi-2021}. Imaginary soft phonon modes in cubic GeTe are also well described by the MLIP, which is essential in capturing the phase transition~\cite{wang-2021,dangic-2021}. Phonon dispersions were computed using the finite displacement method implemented in \textsc{Phonopy}~\cite{togo-2015a} with $3 \times 3 \times 3$ and $5 \times 5 \times 2$ supercells of the conventional unit cells for cubic and rhombohedral phases, respectively. For the DFT calculations, we used the same settings as those used to generate the reference dataset. LO-TO splitting was not included in our calculations as long-range Coulomb interactions tend to be screened by free carriers in real samples~\cite{steigmeier-1970}.

\begin{figure}[t]
    \centering
    \includegraphics[width=0.95\linewidth]{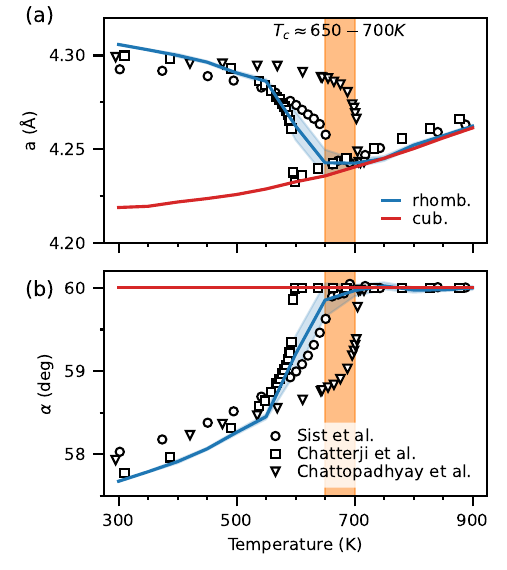}
    \phantomsubfloat{\label{fig:lattice-temp-a}}
    \phantomsubfloat{\label{fig:lattice-temp-angle}}
    \caption{Evolution of \protect\subref{fig:lattice-temp-a} the lattice parameter $a$ and \protect\subref{fig:lattice-temp-angle} the angle $\alpha$ as a function of temperature in the NpT MD simulations of crystalline GeTe, compared against experimental data from Ref.~\cite{sist-2018,chatterji-2015,chattopadhyay-1987}. Simulated lattice parameters in \protect\subref{fig:lattice-temp-a} were shifted by $-0.1~\text{\AA}$.}
    \label{fig:lattice-temp}
\end{figure}

\section{Results and discussion}

\subsection{Finite temperature lattice dynamics}
\label{sec:lattice_dynamics}

We investigated the lattice dynamics of GeTe through MD simulations across the $\alpha \rightarrow \beta$ phase transition with our MLIP. 
For each temperature, rhombohedral and cubic GeTe supercells were first equilibrated for at least 200~ps in the NpT ensemble at ambient pressure with a 2~fs timestep in order to obtain the averaged temperature-dependent structural parameters shown in Fig.~\ref{fig:lattice-temp}. 
The rhombohedral lattice parameter $a$ and angle $\alpha$ reach cubic values for T>650~K, in good agreement with experimental data. Interestingly, below the Curie temperature, the cubic phase remains confined within a local energy minimum and does not undergo transformation into the rhombohedral phase throughout our MD simulations.

\begin{figure}[h!]
    \centering
    \includegraphics[width=0.95\linewidth]{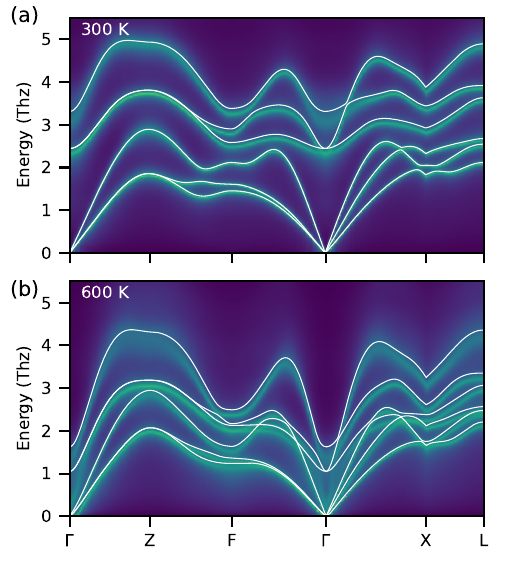}
    \phantomsubfloat{\label{fig:tdep-rhomb-300K}}
    \phantomsubfloat{\label{fig:tdep-rhomb-600K}}
    \caption{Temperature dependent phonon dispersions (white lines) and spectral functions of $\alpha$-GeTe computed using the TDEP method at \protect\subref{fig:tdep-rhomb-300K} 300~K and \protect\subref{fig:tdep-rhomb-600K} 600~K (just before the $R3m \rightarrow Fm\bar{3}m$ phase transition).}
    \label{fig:tdep-rhomb}
\end{figure}

\begin{figure}[h!]
    \centering
    \includegraphics[width=0.95\linewidth]{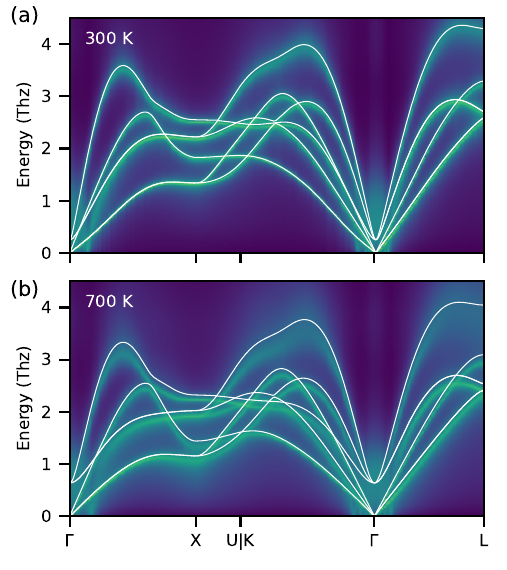}
    \phantomsubfloat{\label{fig:tdep-cub-300K}}
    \phantomsubfloat{\label{fig:tdep-cub-700K}}
    \caption{Temperature dependent phonon dispersions (white lines) and spectral functions of $\beta$-GeTe computed using the TDEP method at \protect\subref{fig:tdep-cub-300K} 300~K and \protect\subref{fig:tdep-cub-700K} 700~K.}
    \label{fig:tdep-cub}
\end{figure}

By employing the temperature-dependent effective-potential (TDEP) method~\cite{hellman-2011,hellman-2013,bottin-2020}, the temperature-dependent interatomic force constants (IFCs) were extracted from configurations sampled from MD trajectories. We used 512-atom rhombohedral or cubic GeTe supercells using the temperature-dependent structural parameters from Fig.~\ref{fig:lattice-temp}. Each structure was equilibrated in the NVT ensemble for 80~ps (thermalization step). Then, snapshots were taken from a 600~ps MD simulation in the microcanonical ensemble for each temperature to extract the IFCs. Details of the TDEP extraction are provided in Appendix~\ref{sec:tdep}. Temperature dependent phonon dispersions and spectral functions obtained using these IFCs are shown in Fig.~\ref{fig:tdep-rhomb} and \ref{fig:tdep-cub} for the $\alpha$ and $\beta$ phases, respectively. 

\begin{figure}[t]
    \centering
    \includegraphics[width=0.95\linewidth]{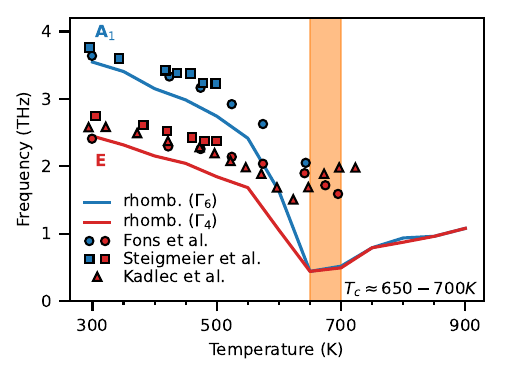}
    \caption{Temperature evolution of A$_1$ and E optical phonon modes computed with the TDEP method and compared against experimental data from Ref.~\cite{fons-2010, kadlec-2011, steigmeier-1970}.}
    \label{fig:phonon-temp}
\end{figure}

In Fig.~\ref{fig:phonon-temp}, the evolution of the longitudinal and transverse optical phonon modes ($\Gamma_6$ and $\Gamma_4$, respectively) is depicted as a function of temperature. The softening of these two modes up to the Curie temperature is consistent with previous theoretical studies~\cite{dangic-2021, wang-2021} and is comparable to experiments~\cite{fons-2010, kadlec-2011, steigmeier-1970}.
Beyond 650~K, the optical phonons merge, indicating the transition to the cubic phase where optical phonons exhibit three-fold degeneracy.

\subsection{Green-Kubo thermal conductivity}

To compute the GK thermal conductivity of both crystalline and amorphous GeTe, MD simulations with the MLIP were performed at different temperatures. The amorphous GeTe structure was generated using a melt-quench process (see Appendix \ref{sec:amorph}). 
Heat flux was computed during MD simulations in the microcanonical ensemble and the ensemble average was carried out over independent trajectories of at least 1~ns after equilibration in the NpT ensemble.

\begin{figure}[h!]
    \centering
    \includegraphics[width=0.95\linewidth]{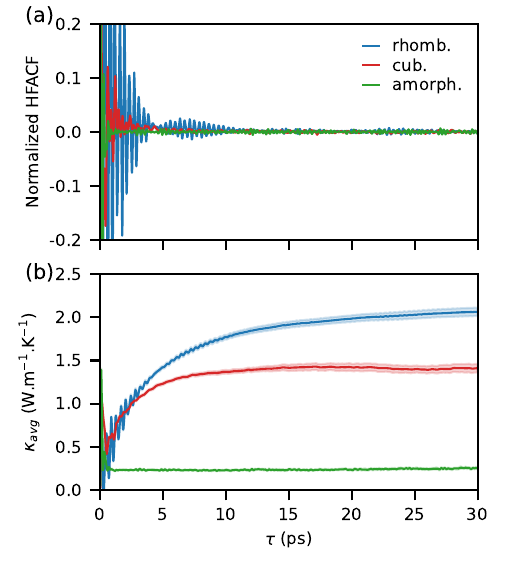}
    \phantomsubfloat{\label{fig:hfacf-300K}}
    \phantomsubfloat{\label{fig:kappa-300K}}
    \caption{\protect\subref{fig:hfacf-300K} Averaged heat flux autocorrelation function (HFACF), normalized by its value at $\tau=0$ and \protect\subref{fig:kappa-300K} lattice thermal conductivity computed using GK theory as a function of correlation time at $T=300$~K for rhombohedral, cubic and amorphous GeTe.}
    \label{fig:md-300K}
\end{figure}

Figure~\ref{fig:hfacf-300K} shows the averaged heat flux autocorrelation function (HFACF) at 300~K as a function of the correlation time $\tau$ 
for the initial GeTe structures in the rhombohedral, cubic, and amorphous phases. The average HFACF features large oscillations before decaying to zero, mainly due to the kinetic contribution of the heat flux (first term in Eq.~\ref{eq:heatflux}). The cumulative time integration of the HFACF yields the thermal conductivities $\kappa_{avg} = 1/3 \sum_{\alpha=1}^{3} \kappa_{\alpha\alpha}$ that are plotted in Fig.~\ref{fig:kappa-300K}.

\begin{figure}[t]
    \centering
    \includegraphics[width=0.95\linewidth]{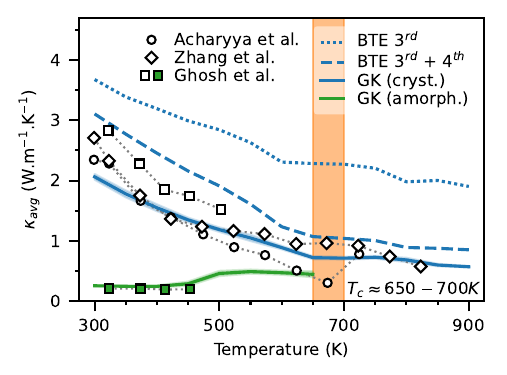}
    \caption{Calculated thermal conductivities using Green-Kubo theory (solid lines) of amorphous and crystalline GeTe. The thermal conductivity computed using direct solution of linearized Boltzmann transport equation are also reported (dotted line for three phonon, dashed line for three and four phonon).
    Experimental data are shown for crystalline GeTe from Ref.~\cite{acharyya-2020, zhang-2021, ghosh-2020}) (empty symbols) and for amorphous GeTe from Refs.~\cite{ghosh-2020} (green symbols).}
    \label{fig:kappa-exp}
\end{figure}

After testing the convergence with respect to system size (see Appendix~\ref{sec:gk-convergence}), we employed supercells containing 360 and 512 atoms for crystalline and amorphous structures, respectively. Figure~\ref{fig:kappa-exp} shows the computed lattice thermal conductivities as a function of temperature for both amorphous and crystalline GeTe together with existing experimental data from Refs.~\cite{acharyya-2020, zhang-2021, ghosh-2020}.
Comparing against experiments is challenging because experimental values of lattice thermal conductivities of crystalline GeTe show a large dispersion. This variability can be attributed to two main reasons.
First, thermal conductivity comprises a lattice contribution and an electronic contribution. Therefore, experimental lattice thermal conductivity is an indirect measurement, which is obtained by removing the electronic contribution from the measured thermal conductivity, typically evaluated using the Wiedmann-Franz law that introduces an additional approximation from the Lorenz number. 
Second, the quality of the samples varies, leading to variations in thermal conductivity measurements. Despite the significant experimental variations mentioned above, the calculated GK thermal conductivity values are found to fall within the range of experimental values.

The GK lattice thermal conductivity for the amorphous phase (solid green line) is in excellent agreement with the experimental data of Ref.~\cite{ghosh-2020} (green squares). 
This can be regarded as a direct comparison with the experiment since the electronic contribution to the thermal conductivity was found to be negligible in amorphous GeTe~\cite{nath-1974}.
A previous study obtained a similar value of $0.27\pm0.05$~W$\cdot$m$^{-1}\cdot$K$^{-1}$ at 300~K from GK simulations with a Behler-Parrinello-type MLIP~\cite{sosso-2012}.
The predicted thermal conductivity for amorphous GeTe is constant until $\sim 450$~K. It then starts to increase, indicating a transition to a crystalline phase, as evidenced by the evolution of the radial distribution function (see Fig.~\ref{fig:rdf}.) and consistent with the amorphous-crystalline phase transition temperature observed experimentally~\cite{ghosh-2020}.

\subsection{BTE thermal conductivity}

To compute the thermal conductivity using the BTE, we employed the FourPhonon code~\cite{han-2022} with the TDEP IFCs up to fourth order and an $11 \times 11 \times 11$ $q$-mesh. 
This allows a direct comparison between GK and BTE methods, as both calculations were conducted under the same conditions, utilizing identical interatomic potentials and temperatures; the only difference lies in the thermal transport formalism employed.
BTE including only three-phonon processes overestimates the thermal conductivity by about 1.8~W$\cdot$m$^{-1} \cdot$K$^{-1}$, which is about twice the GK result at 300~K, and about three times that at 900~K. 
Including four-phonon processes significantly improves BTE results. The BTE thermal conductivity approaches that of GK, especially at high temperatures in the cubic phase. However, the discrepancy at 300~K remains about $50\%$.
Such overestimation indicates that higher-order IFCs are mandatory in BTE to capture the strong anharmonicity. 
Furthermore, the computational expenses escalate considerably when incorporating four-phonon processes in BTE. Consequently, the GK method emerges as a viable approach for assessing the lattice thermal conductivity of materials exhibiting strong anharmonicity.
We observed differences when comparing our results to previous BTE calculations reported in the literature~\cite{dangic-2021,ghosh-2020,xia-2018}. These discrepancies are discussed in Appendix~\ref{sec:previous_works}.

\section{Conclusion}

In conclusion, we developed an equivariant graph neural network interatomic potential to study thermal transport in both amorphous and crystalline GeTe. The potential describes GeTe at a near-\textit{ab~initio} level of accuracy for the rhombohedral, cubic and amorphous phases with a single model.
Furthermore, it successfully captures phase transitions with transition temperatures in good agreement with experimental data.
When coupled with Green-Kubo theory, it can determine the lattice thermal conductivity not only for strongly anharmonic crystals, but also for the amorphous phase.

\begin{acknowledgments}
We thank F. Bottin and J. Bouchet for discussions about TDEP calculations. This work was performed using HPC/AI resources from GENCI–IDRIS (Grant No. 2022-A0110911995) and was partially funded by the European commission through ECSEL-IA 101007321 project StorAIge and the French IPCEI program.
\end{acknowledgments}

\appendix

\section{Machine Learning Interatomic Potential generation}
\label{sec:mlip}

\subsection{Training}
The equivariant graph neural network (GNN) model was trained using the \textsc{Allegro} code~\cite{musaelian-2023} on an NVIDIA A100 GPU with \texttt{float32} precision. The dataset contained 6000 reference configurations, randomly split into 5000, 500 and 500 configurations for the training, validation and test sets, respectively~\cite{lee-2024}. The training was done using a mean squared error (MSE) loss function based on a weighted sum of energy, forces and stress loss terms:
\begin{widetext}
\begin{equation}
\label{eq:loss}
\mathcal{L} = \dfrac{\lambda_E}{B} \sum_b^B \left( \dfrac{\hat{E}_b - E_b}{N} \right)^2 + \dfrac{\lambda_F}{3BN} \sum_{i=1}^{BN} \sum_{\alpha=1}^{3}\left( \hat{F}_{i,\alpha} - F_{i,\alpha} \right)^2 + \dfrac{\lambda_\sigma}{9B} \sum_{b}^{B} \sum_{\alpha=1}^{3} \sum_{\beta=1}^{3} \left( \hat\sigma_{b,\alpha\beta} - \sigma_{b,\alpha\beta}\right)^2
\end{equation}
\end{widetext}
where $B$ is the batch size, $N$ is the number of atoms, $E_b$ is the DFT energy, $\hat{E}_b$ is the predicted energy, $F_{i,\alpha}$ is the DFT force of atom $i$ along direction $\alpha$, $\hat{F}_{i,\alpha}$ is the predicted force of atom $i$ along direction $\alpha$, $\sigma_{b,\alpha\beta}$ is the DFT stress tensor $\alpha\beta$ component and $\hat\sigma_{b,\alpha\beta}$ is the predicted stress tensor $\alpha\beta$ component. We used the scaling factors $\lambda_E = 1$, $\lambda_F = 1$ and $\lambda_\sigma = 10$ (with energy, force and stress units in eV, $\text{eV}\cdot\text{\AA}^{-1}$ and $\text{eV}\cdot\text{\AA}^{-3}$, respectively).

\subsection{Hyperparameters}

The scalar latent features were generated via a projection onto eight trainable Bessel functions and the associated two-body multi layer perceptron (MLP) consisted of four hidden layers of dimensions [128, 256, 512, 1024] with SiLU nonlinearities~\cite{hendrycks_gaussian_2023}. 
A cutoff function with a cutoff radius of 6.5~\AA{} and a polynomial envelope function with $p=6$ was used.
For the equivariant features, a maximum rotation order $l_{max} = 2$ with 16 features of even parity was used.
The model contained one interaction layer, where the latent MLP contained three hidden layers of dimensions [1024, 1024, 1024] with SiLU nonlinearities~\cite{hendrycks_gaussian_2023} and the embedding MLP for the tensor product operation contained a single layer with no nonlinearity.
The final output MLP was a single layer with 128 nodes and no nonlinearity.

Training was performed with a batch size of 5 and the Adam optimizer was used with an initial learning rate of 0.01, decaying by a factor of 0.5 whenever the validation loss had not seen an improvement for 20 epochs.

\section{TDEP}
\label{sec:tdep}

\begin{figure}[t]
    \centering
    \includegraphics[width=0.95\linewidth]{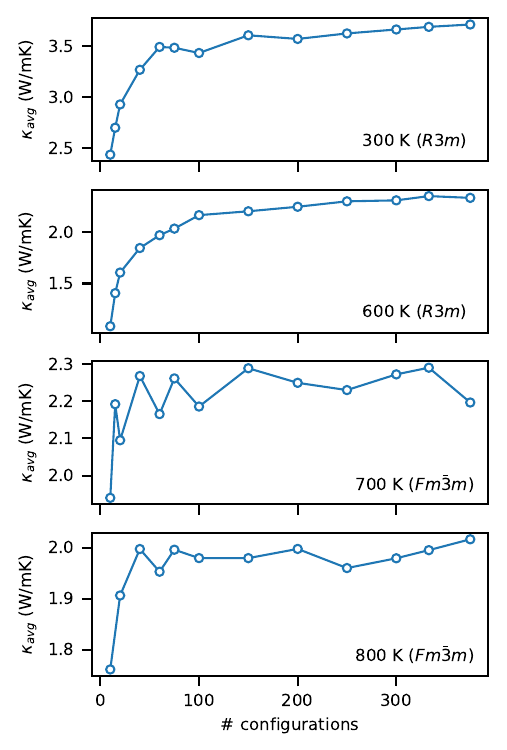}
    \phantomsubfloat{\label{fig:tdep-nconfig-300K}}
    \phantomsubfloat{\label{fig:tdep-nconfig-600K}}
    \caption{Thermal conductivity of GeTe as a function of the number of configurations used to extract second and third order force constants using the TDEP method.}
    \label{fig:tdep-nconfig}
\end{figure}

For TDEP calculations, we used cutoff radii of 12, 8 and 5~$\text{\AA}$ to extract second, third and fourth-order IFCs, respectively. We verified that the employed cutoff radii lead to converged results. We also tested the convergence of the thermal conductivity values with respect to the number of configurations for different temperature, as shown in  Fig.~\ref{fig:tdep-rhomb}. We used 300 configurations to extract the IFCs used to compute phonon dispersions and thermal conductivities presented in Sec.~\ref{sec:lattice_dynamics}.

In addition, for the rhombohedral phase, we used equilibrium atomic positions computed as an average over all MD snapshots to avoid the spurious soft LO phonon mode at $\Gamma$. For the cubic phase we used ideal atomic positions.

\section{Amorphous structure}
\label{sec:amorph}

\begin{figure}[h]
    \centering
    \includegraphics[width=0.68\linewidth]{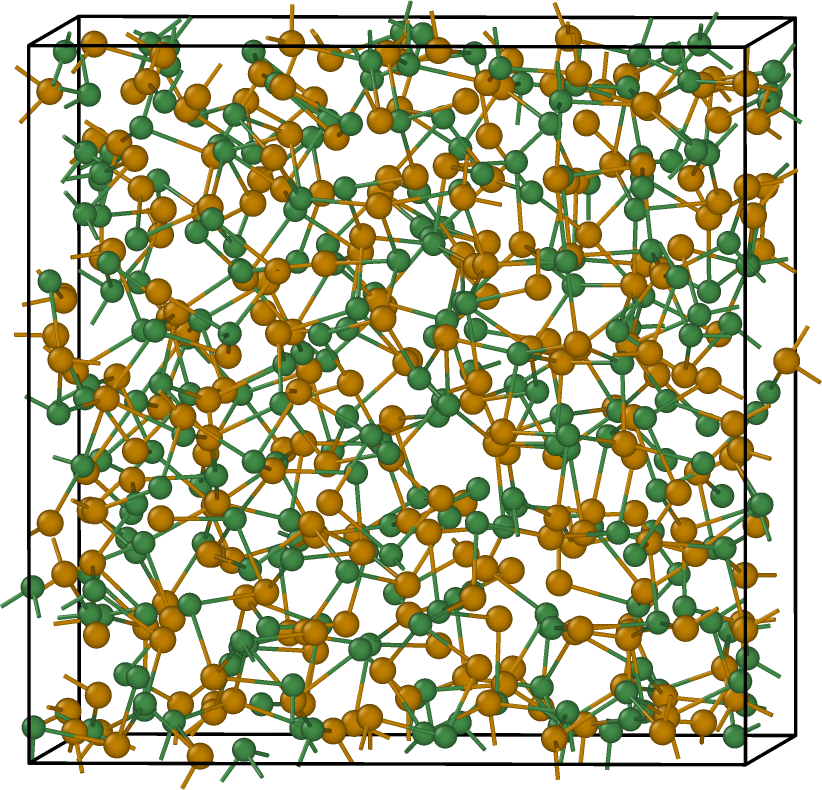}
    \caption{Amorphous GeTe structure generated by melt-quench MD simulation.}
    \label{fig:amorph}
\end{figure}

The amorphous structure was generated by a melt-quench (MQ) MD simulation using our GeTe MLIP. We used a 512-atom cubic GeTe supercell at the experimental density (0.03327~atoms/$\text{\AA}^3$~\cite{ghezzi-2011}) as the starting structure. The structure was first melted at 1800~K for 200~ps, then quenched to 300~K over 200~ps in the NVT ensemble. Finally, the structure was equilibrated in the NpT ensemble at 300~K for 1~ns. The density of the final structure was 0.03328~atoms/$\text{\AA}^3$, very close to the experimental value. The amorphous structure is shown in Fig.~\ref{fig:amorph}.

\begin{figure}[h]
    \centering
    \includegraphics[width=0.9\linewidth]{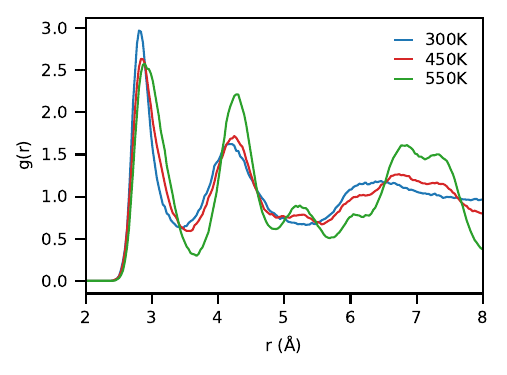}
    \caption{Radial distribution function of the amorphous structure after equilibration at different temperatures.}
    \label{fig:rdf}
\end{figure}

For the lattice thermal conductivity calculations, the amorphous structure was equilibrated for 0.8~ns in the NpT ensemble at different temperatures before computing the heat flux in the microcanonical ensemble. We computed the time-averaged radial distribution function (RDF) $g(r)$ at the end of the NpT simulation for 10~ps (see Fig.~\ref{fig:rdf}). Between 300~K and 450~K, the RDFs are very similar and show no long range order beyond $6~\text{\AA}$. In contrast, at 550~K the RDF displays some peaks beyond $6~\text{\AA}$, which suggests a crystallization of the structure. 

\section{Convergence of Green-Kubo simulations}
\label{sec:gk-convergence}

\begin{figure}[t]
    \centering
    \includegraphics[width=0.95\linewidth]{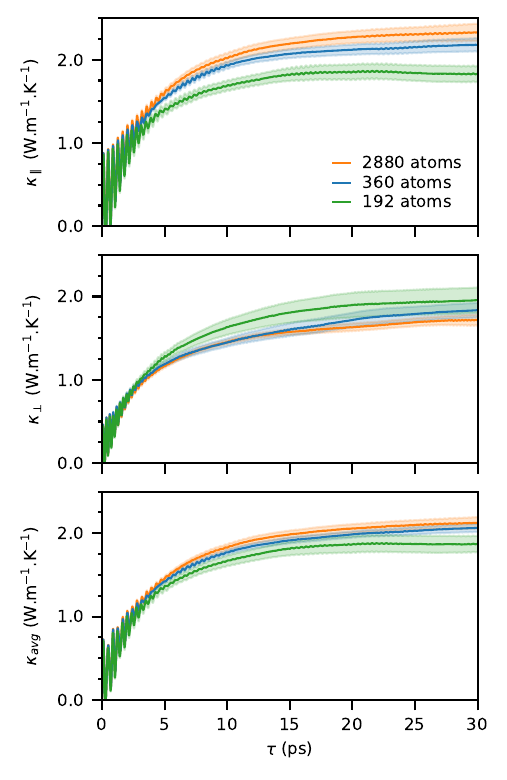}
    \caption{Lattice thermal conductivity of rhombohedral GeTe as a function of correlation time computed at 300~K using GK method for different system sizes.}
    \label{fig:gk-size-rhomb}
\end{figure}

Green-Kubo calculations require large system sizes and long simulation times in order to obtain a converged thermal conductivity. We computed the lattice thermal conductivity of the rhombohedral phase using supercells with 192, 360, and 2880 atoms. For each system size, we performed 10 independent MD simulations with different initial velocities. The averaged thermal conductivity over the different MD trajectories is shown in Fig.~\ref{fig:gk-size-rhomb}, along with the standard error (shaded region). The thermal conductivity computed for the 2880-atom supercell exhibits larger standard errors compared to 360- and 192-atom supercells due to the shorter simulation times used. The lattice thermal conductivities are extracted for correlation times where the averaged thermal conductivity reaches a plateau (i.e., typically for $\tau > 25$~ps). 

\section{Comparison with previous theoretical works}
\label{sec:previous_works}

\begin{figure}[t]
    \centering
    \includegraphics[width=0.95\linewidth]{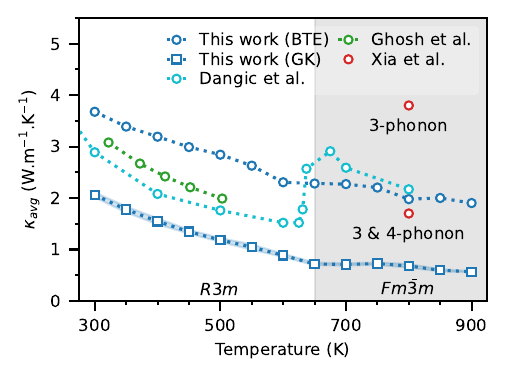}
    \caption{Lattice thermal conductivity of crystalline GeTe (rhombohedral $R3m$ at the left, and cubic $Fm\bar{3}m$ phase at the right) as a function of temperature calculated by the Green-Kubo (GK, squares) or the Boltzmann transport equation (BTE, circles) methods and compared with previous theoretical works of Ref.~\cite{dangic-2021,ghosh-2020,xia-2018}, all solving the BTE.}
    \label{fig:kappa-comp}
\end{figure}

Figure~\ref{fig:kappa-comp} shows the comparison of calculated lattice thermal conductivity of crystalline GeTe using the BTE (circles) or GK (squares) method, compared to previous theoretical works from Refs.~\cite{dangic-2021,ghosh-2020,xia-2018}. It should be noted that for $T \lesssim 650$~K, GeTe is in the rhombohedral ($R3m$) phase, while for $T \gtrsim 650$~K, GeTe is in the cubic ($Fm\bar{3}m$) phase. In our calculations, we used the structural parameters obtained from the NpT MD simulations (which captures the phase transition at $T \approx 650$~K). The BTE calculations shown in Fig.~\ref{fig:kappa-comp} include three-phonon processes using IFCs extracted with the TDEP method. BTE thermal conductivities including three- and four-phonon processes are shown in Fig.~\ref{fig:kappa-exp}.

\begin{figure}[th!]
    \centering
    \includegraphics[width=0.95\linewidth]{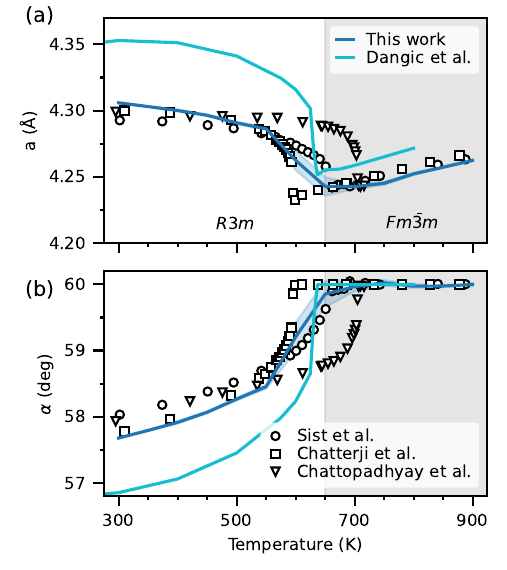}
    \phantomsubfloat{\label{fig:lattice-comp-a}}
    \phantomsubfloat{\label{fig:lattice-comp-angle}}
    \caption{Evolution of \protect\subref{fig:lattice-comp-a} the lattice parameter $a$ and \protect\subref{fig:lattice-comp-angle} the angle $\alpha$ as a function of temperature in the NpT MD simulations of crystalline GeTe (starting with a $R3m$ supercell), compared against simulations from Ref.~\cite{dangic-2018} and experimental data from Ref.~\cite{sist-2018,chatterji-2015,chattopadhyay-1987}. Simulated lattice parameters in \protect\subref{fig:lattice-comp-a} were shifted by $-0.1~\text{\AA}$.}
    \label{fig:lattice-comp}
\end{figure}

A similar approach was employed in Ref.~\cite{dangic-2021} (light blue curve in Fig.~\ref{fig:kappa-comp}) but the reported values are significantly different compared to our results for the rhombohedral phase. The primary source of discrepancy seems to be related to the convergence with respect to the number of configurations used to extract second and third order TDEP IFCs. In Ref.~\cite{dangic-2021}, the authors used only 24 configurations. According to our convergence test (see Fig.~\ref{fig:tdep-nconfig}), such a small number of configurations results in strongly underestimated thermal conductivity values. Another source of discrepancy lies in the temperature evolution of the structural parameters (see Fig.~\ref{fig:lattice-comp}) which may also lead to some differences in the computed lattice thermal conductivities. For the cubic phase ($T \geq 700$~K), our results show a good agreement with those of Ref.~\cite{dangic-2021}, presumably because less configurations are required to achieve convergence compared to the rhombohedral symmetry.

In Ref.~\cite{ghosh-2020} (green curve in Fig.~\ref{fig:kappa-comp}), the authors used  0~K IFCs and structural parameters of $\alpha$-GeTe to compute the BTE thermal conductivity. Therefore, their results exhibit a different temperature evolution.

Finally, in Ref.~\cite{xia-2018}, temperature-dependent IFCs were computed for the $\beta$ phase of GeTe up to fourth order (i.e. three- and four-phonon scattering processes) and BTE was solved under the relaxation time approximation at 800~K. Their thermal conductivity value  considering only three-phonon scattering is significantly larger compared to the one from our work or Ref.~\cite{dangic-2021}. Including four-phonon scattering reduces the thermal conductivity by the same factor as in our work (reduction by a factor 2.23).

\newpage

\providecommand{\noopsort}[1]{}\providecommand{\singleletter}[1]{#1}%

\end{document}